\documentclass[preprint,showpacs,preprintnumbers,amsmath,amssymb]{revtex4}
\usepackage{graphicx}
\usepackage{dcolumn}
\usepackage{bm}

\begin{document}

\title{Long-range correlation and multifractality in Bach's Inventions pitches}

\author{G. R. Jafari, P. Pedram}
 \affiliation{Department of physics, Shahid Beheshti University,
Evin, Tehran 19839, Iran \\ Department of nano-science, IPM, P. O.
Box 19395-5531, Tehran, Iran}
\author{L. Hedayatifar}
\affiliation{Department of physics, Yazd University, Yazd, Iran }
\date{\today}

\begin{abstract}
We show that it can be considered some of Bach pitches series as a
stochastic process with  scaling behavior. Using multifractal
deterend fluctuation analysis (MF-DFA) method, frequency series of
Bach pitches have been analyzed. In this view we find same second
moment exponents (after double profiling) in ranges ($1.7-1.8$) in
his works. Comparing MF-DFA results of original series to those for
shuffled and surrogate series we can distinguish multifractality due
to long-range correlations and a broad probability density function.
Finally we determine the scaling exponents and singularity spectrum.
We conclude fat tail has more effect in its multifractality nature
than long-range correlations.
\end{abstract}


\maketitle

{\it Pythagoras knew it, but Bach demonstrated it: without
mathematics there is no music} \cite{Noralv}.

\section{\label{sec:level1}Introduction}

Many people think that, mathematics and music have some vague sort
of affinity, but most often supposed relationship between  two
fields turns out to be in details that are not central to either.
The mathematical proportions in musical works are hidden from
listener from old days. Thus it is forced to make use of
interpretative techniques in order to search for them, which is
problematic from a methodological point of view. Some music
historians have very little time for numerology. In addition to
mathematics being seen as numerical symbolism, music is closely
linked to absolute physical entities, such as frequency and relation
between intervals (an interval is space between two notes). Already
in antiquity this was seen as  natural or cosmic premise on which
music relied. It is illustrated fact that not just musical notation,
but also  relationship between music and time has something to do
with mathematics and with one of the most significant
transformations in music history. Complexity of music can be
especially attracting scientific interest. Among great variety of
complex and disordered systems most of  music parameters such as
frequency and pitch (Pitch is the sound frequency of any given
note.) \cite{lyan, Heather, Jafari, Gonzalez, Shi}, Amplitude or
Dynamics (Dynamics are the changes in volume during a musical
piece.) \cite{Diodati, Jean}, Intervals (Intervals are the distances
between notes in the musical scale.), Rhythm (Rhythm is the
structure of the placement of notes in musical time.) can be
consider as stochastic processes. Also, some authors try to
clustering the music \cite{Rudi1, Rudi2}.

In all technicalities, music can be composed of notes. A note is a
sign used in music to represent the relative duration and pitch of
sound. In traditional music theory pitch classes are represented by
the first seven letters of the Latin alphabet (A, B, C, D, E, F, and
G) or (\textit{Do - Re - Mi - Fa - Sol - La - Si}), in Italian
notation. Each note is assigned a specific vertical position on a
staff. Since the physical causes of music are vibrations of
mechanical systems, their frequencies are often measured in hertz
(Hz). These frequencies are mathematically related to each other,
and are defined around the central note. The current "standard
pitch" for this note is 440 Hz. Any note is exactly an integer
number of half-steps away from central note. Let this distance be
denoted n. Then, the desired frequency is given by
\begin{equation}
\nu = 440 \times 2^{n/12} \ \mbox{Hz}.
\end{equation}

In this paper we would like to characterize complex behavior of
frequency of note signal of Bach's Inventions and Sinfonias through
computation of signal parameters scaling exponents, which quantifies
correlation exponents and multifractality of signal. Inventions and
Sinfonias are collection of short pieces which Bach wrote for
musical education of Bach young pupils. These are among finest
examples of artistic gems ever written for this purpose, and
probably because of this, they became very popular among Bach's
pupils and others ever since they were written. Inventions and
Sinfonias contain two and three music voices respectively. The
number of the data in frequency series is dependent to the piece and
will obtain from one voice which contains 1000 notes in average.
Because of non-stationary nature of frequency of music series, and
due to finiteness of available data sample, we should apply a
methods which are insensitive to non-stationarities, like trends. In
order to separate trends from correlations we need to eliminate
trends in our frequency data. Several methods are used effectively
for this purpose: detrended fluctuation analysis (DFA)
\cite{Peng94}, rescaled range analysis (R/S) \cite{hurst65} and
wavelet techniques (WT) \cite{wtmm}.

\begin{figure}
\begin{center}
\includegraphics[width=12cm]{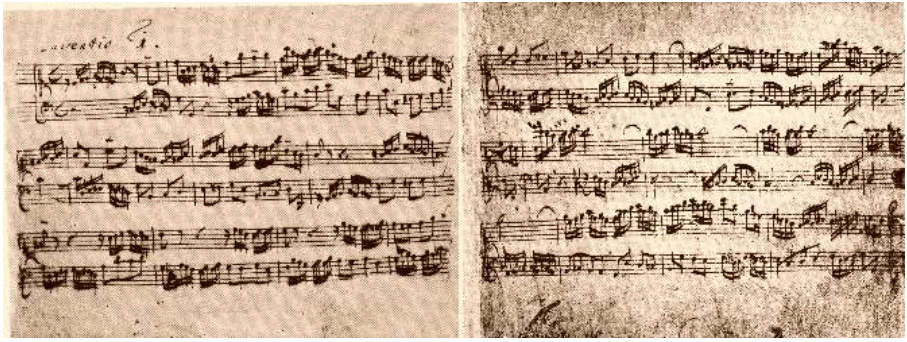}
\end{center}
\caption{Typical sheet music, Invention no.1 by Bach.}
\label{FIG:nsvst1}
\end{figure}

We use MF-DFA method for analysis and eliminating  trend from data
set. This method are the modified version of DFA method to detect
multifractal properties of time series. DFA method introduced by
Peng et al. \cite{Peng94} has became a widely used technique for
the determination of (mono-) fractal scaling properties and the
detection of long-range correlations in noisy, non-stationary time
series \cite{murad,physa,kunhu,kunhu1}. It has successfully been
applied to diverse fields such as DNA sequences \cite{Peng94,dns},
heart rate dynamics \cite{herz,Peng95,PRL00}, neuron spiking
\cite{neuron}, human gait \cite{gait}, wind speed,
\cite{Govindan}, long-time weather records \cite{wetter}, cloud
structure \cite{cloud}, geology \cite{malamudjstatlaninfer1999},
ethnology \cite{Alados2000}, economical time series
\cite{economics}, and solid state physics \cite{fest}. One reason
to employ  DFA method is to avoid spurious detection of
correlations that are artefacts of non-stationarity in time
series.

The focus of present paper is on intriguing statistical properties
and multifractal nature of frequency series. In Particular, Figures
1 and 2 show the score of Invention no.1 and the frequency
fluctuation of Invention no. 1 and Sinfoina no. 1, respectively. In
general, two different types of multifractality in frequency series
can be distinguished: (i) Multifractality due to a fatness of
probability density function (PDF) of  time series. In this case
multifractality cannot be removed by shuffling the series. (ii)
Multifractality due to different long-range correlations in small
and large scale fluctuations. In this case  data may have a PDF with
finite moments, e.g. a Gaussian distribution. Thus corresponding
shuffled time series will exhibit mono-fractal scaling, since all
long-range correlations are destroyed by shuffling procedure. If
both kinds of multifractality are present, shuffled series will show
weaker multifractality than original series.

The paper is organized as follows: In section II we describe MF-DFA
methods in detail and show that, scaling exponents determined by
MF-DFA method are identical to those obtained by standard
multifractal formalism based on partition functions. In section III,
we analysis of frequency series of Bach's Inventions also examine
source of multifractality in frequency data by comparison MF-DFA
results for remaining data set to those obtained by MF-DFA for
shuffled and surrogate series. section IV
closes with a conclusion.\\

\section{Multifractal Detrended Fluctuation Analysis}
The simplest type of multifractal analysis is based upon standard
partition function multifractal formalism, which has been developed
for  multifractal characterization of normalized, stationary
measurements \cite{feder88,barabasi,peitgen,bacry01}. Unfortunately,
this standard formalism does not give us correct results for
non-stationary time series that are affected by trends or those
which cannot be normalized. In the early 1990s an improved
multifractal formalism wavelet transform modulus maxima (WTMM)
method \cite{wtmm}, has been developed. This method is based on
wavelet analysis and involves tracing the maxima lines in continuous
wavelet transform over all scales. Other method like, multifractal
detrended fluctuation analysis (MF-DFA), is based on identification
of scaling of $q$th-order moment depending on signal length, and it
 is generalization of  standard DFA method in which
$q=2$. In contrast to WTMM method  MF-DFA does not require modulus
maxima procedure, and hence does not require more effort in
programming and computing time than conventional DFA.
 On the other hand,
one should find correct scaling behavior of fluctuations, from
experimental data which often affected by non-stationary like
trends. This have to be well distinguished from intrinsic
fluctuations of system.
 In
addition often in collected data
 we do not know  reasons, or even worse scales,
 for underlying trends, and
  also, usually  available record data is small.
So for reliable detection of correlations, it is essential to
distinguish trends for intrinsic fluctuations from collected data.
Hurst rescaled-range analysis \cite{hurst65} and other
non-detrending methods work well when records are long and do not
involve trends, otherwise it might give wrong results.
 DFA is a well established method for determining
 scaling behavior of noisy data, where data presence of trends and we
don't know their origin and shape
\cite{Peng94,Peng95,fano,allan,buldy95}.

\begin{figure}[t]
\includegraphics[width=9cm]{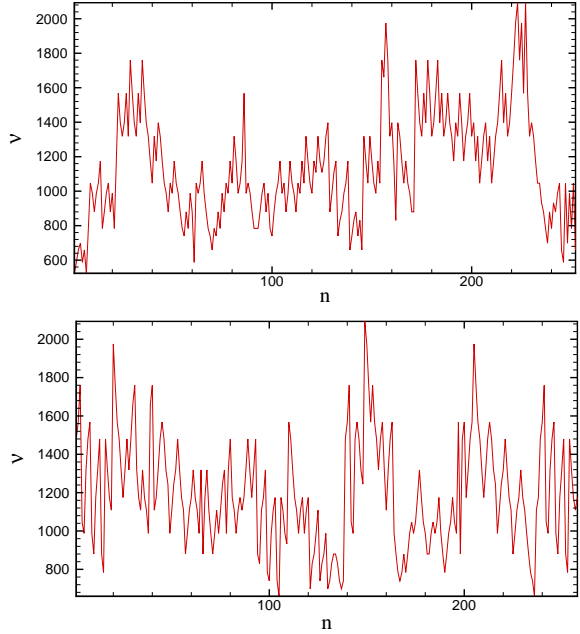}
\caption{ Up to down: typical frequency series of Invention and
Sinfonias no.1 by Bach.} \label{FIG:nsvst2}
\end{figure}

\subsection{Description of  MF-DFA method}

Modified multifractal DFA (MF-DFA) procedure consists of five
steps. The first three steps are essentially identical to
conventional DFA procedure (see e.g.
\cite{Peng94,murad,physa,kunhu,kunhu1}). Suppose that $x_k$ is a
series of length $N$, and it is of compact support, i.e. $x_k = 0$
for an insignificant fraction of the values only.

\noindent $\bullet$ {\it Step 1}: Determine the ``profile''
\begin{equation} Y(i) \equiv \sum_{k=1}^i \left[ x_k - \langle x
\rangle \right], \qquad i=1,\ldots,N. \label{profile}
\end{equation}
Subtraction of the mean $\langle x \rangle$ from $x_k$ is not
compulsory, since it would be eliminated by later detrending in
third step.

\noindent $\bullet$ {\it Step 2}: Divide  profile $Y(i)$ into $N_s
\equiv {\rm int}(N/s)$ non overlapping segments of equal lengths
$s$. Since  length $N$ of  series is often not a multiple of
considered time scale $s$, a short part at the end of profile may
remain.  In order not to disregard this part of series, same
procedure should be repeated starting from the opposite end.
Thereby, $2 N_s$ segments are obtained altogether.

\noindent $\bullet$ {\it Step 3}: Calculate  local trend for each
of $2 N_s$ segments by a least-square fit of series. Then
determine the variance
\begin{equation} F^2(s,\nu) \equiv {1 \over s} \sum_{i=1}^{s}
\left\{ Y[(\nu-1) s + i] - y_{\nu}(i) \right\}^2, \label{fsdef}
\end{equation}
for each segment  $\nu = 1, \ldots, N_s$ and
\begin{equation} F^2(s,\nu) \equiv {1 \over s} \sum_{i=1}^{s}
\left\{ Y[N - (\nu-N_s) s + i] - y_{\nu}(i) \right\}^2,
\label{fsdef2}
\end{equation}
for $\nu = N_s+1, \ldots, 2 N_s$.  Where, $y_{\nu}(i)$ is fitted
polynomial in segment $\nu$.  Linear, quadratic, cubic, or higher
order polynomials can be used in the fitting procedure
(conventionally called DFA1, DFA2, DFA3, $\ldots$, DFA$m$)
\cite{Peng94,PRL00}. Since detrending of time series is done by
subtraction of fited polynomial  from profile, different order DFA
differ in their capability of eliminating trends in series.  In
(MF-)DFA$m$ trend of order $m$ in  profile (and equivalently, order
$m - 1$ in original series) are eliminated.  Thus a comparison of
results for different orders of DFA allows one to estimate type of
the polynomial trend in time series \cite{physa,kunhu}.

\noindent $\bullet$ {\it Step 4}: Average over all segments to
obtain $q$-th order fluctuation function, defined by:
\begin{equation} F_q(s) \equiv \left\{ {1 \over 2 N_s}
\sum_{\nu=1}^{2 N_s} \left[ F^2(s,\nu) \right]^{q/2}
\right\}^{1/q}, \label{fdef}\end{equation}
where, in general, variable $q$ can take any real value except
zero.   $q=2$, standard DFA procedure is retrieved. Generally we
are interested to know  how  generalized $q$ dependent fluctuation
functions $F_q(s)$ depend on  time scale $s$ for different values
of $q$.  Hence, we must repeat steps 2, 3 and 4 for several time
scales $s$.  It is apparent that $F_q(s)$ will increase with
increasing $s$.  Of course, $F_q(s)$ depends on  DFA order $m$. By
construction, $F_q(s)$ is only defined for $s \ge m+2$.

\noindent $\bullet$ {\it Step 5}: Determine scaling behavior of
fluctuation functions by analyzing log-log plots of $F_q(s)$
versus $s$ for each value of $q$. If series $x_i$ are long-range
power law correlated, then $F_q(s)$ , for large values of $s$,
increases as a power-law i.e.,
\begin{equation} F_q(s) \sim s^{h(q)} \label{Hq}. \end{equation}
In general, exponent $h(q)$ may depend on $q$. For stationary time
series such as, fractional Gaussian noise (fGn), $Y(i)$ in eq.
\ref{profile}, will have a fractional Brownian motion (fBm) signal,
so, $0<h(q=2)<1.0$. The exponent $h(2)$ is identical to well known
Hurst exponent $H$ \cite{Peng94,murad,feder88}. Also for
non-stationary  signal, such as fBm noise, $Y(i)$ in eq.
\ref{profile}, will be a sum of fBm signal, so  corresponding
scaling exponent of $F_q(s)$ is identified by $h(q=2)>1.0$
\cite{Peng94,eke02}. For monofractal time series, $h(q)$ is
independent of $q$, since scaling behavior of  variance $F^2(s,\nu)$
is identical for all segments $\nu$, and averaging procedure in
eq.(\ref{fdef}) will just give this identical scaling behavior for
all values of $q$. If we consider positive values of $q$, the
segments $\nu$ with large variance $F^2(s,\nu)$ (i.e. large
deviation from the corresponding fit) will dominate  average
$F_q(s)$.  Thus, for positive values of $q$, $h(q)$ describes
scaling behavior of segments with large fluctuations. For negative
values of $q$, segments $\nu$ with small variance $F^2(s,\nu)$ will
dominate average $F_q(s)$. Hence, for negative values of $q$, $h(q)$
describes  scaling behavior of segments with small fluctuations.

\subsection{Relation to standard multifractal analysis}

For a stationary, normalized series  multifractal scaling exponent
$h(q)$ defined in eq.(\ref{Hq}) are directly related to scaling
exponent $\tau(q)$ defined by standard partition function based on
multifractal formalism as shown below. Suppose that series $x_k$
of length $N$ is a stationary, normalized sequence, then
detrending procedure in step 3 of MF-DFA method is not required,
since no trend has to be eliminated. Thus,  DFA can be replaced by
standard fluctuation analysis (FA), which is identical to DFA
except  definition of variance, which is simplified
 for each segment  $\nu = 1, \ldots, N_s$. In step 3 eq.(\ref{fsdef})
now becomes :
\begin{equation} F_{\rm FA}^2(s,\nu) \equiv [Y(\nu s) - Y((\nu-1) s)]^2.
\label{FAfsdef} \end{equation}
Inserting this simplified definition into eq.(\ref{fdef}) and
using eq.(\ref{Hq}), we obtain
\begin{equation} \left\{ {1 \over 2 N_s} \sum_{\nu=1}^{2 N_s}
\vert Y(\nu s) - Y((\nu-1) s) \vert^q \right\}^{1/q} \sim
s^{h(q)}. \label{FAfHq} \end{equation}
For simplicity we can assume that  length $N$ of series is an
integer multiple of scale $s$, obtaining $N_s = N/s$ and therefore
\begin{equation} \sum_{\nu=1}^{N/s} \vert Y(\nu s) - Y((\nu-1) s)
\vert^q \sim s^{q h(q) - 1}. \label{MFA} \end{equation}
This corresponds to the multifractal formalism used e.g. in
\cite{barabasi,bacry01}. In fact, a hierarchy of exponents $H_q$
similar to our $h(q)$ has been introduced based on eq.(\ref{MFA})
in \cite{barabasi}. In order to relate this to  standard textbook
box counting formalism \cite{feder88,peitgen}, we employ
definition of profile in eq.(\ref{profile}). It is evident that
term $Y(\nu s) - Y((\nu-1) s)$ in eq.(\ref{MFA}) is identical to
sum of numbers $x_k$ within each segment $\nu$ of size $s$. This
sum is known as  box probability $p_s(\nu)$ in standard
multifractal formalism for normalized series $x_k$,
\begin{equation} p_s(\nu) \equiv \sum_{k=(\nu-1) s +1}^{\nu s} x_k =
Y(\nu s) - Y((\nu-1) s).  \label{boxprob} \end{equation}
The scaling exponent $\tau(q)$ is usually defined via partition
function $Z_q(s)$,
\begin{equation} Z_q(s) \equiv \sum_{\nu=1}^{N/s} \vert p_s(\nu)
\vert^q \sim s^{\tau(q)}, \label{Zq} \end{equation}
where $q$ is a real as in MF-DFA method, discussed above. Using
eq.(\ref{boxprob}) we see that eq.(\ref{Zq}) is identical to
eq.(\ref{MFA}), and one can obtain analytical relation between
two sets of multifractal scaling exponents,
\begin{equation} \tau(q) = q h(q) - 1. \label{tauH} \end{equation}
Thus, we see that $h(q)$ defined in eq.(\ref{Hq}) for MF-DFA is
directly related to classical multifractal scaling exponents
$\tau(q)$.  Note that, $h(q)$ is different from generalized
multifractal dimension
\begin{equation} D(q) \equiv {\tau(q) \over q-1} =
{q h(q)-1 \over q-1}, \label{Dq} \end{equation} that are used
instead of $\tau(q)$ in some papers.  In this case, while $h(q)$
is independent of $q$ for a monofractal time series, $D(q)$
depends on $q$.
 Another way to characterize a multifractal series is looking to
singularity spectrum $f(\alpha)$, which is related to $\tau(q)$
via a Legendre transform \cite{feder88,peitgen},
\begin{equation} \alpha = \tau'(q) \quad {\rm and} \quad
f(\alpha) = q \alpha - \tau(q). \label{Legendre} \end{equation}
Here, $\alpha$ is the singularity strength or H\"older exponent,
while $f(\alpha)$ denotes  dimension of subset of  series that is
characterized by $\alpha$. Using eq.(\ref{tauH}), we can directly
relate $\alpha$ and  $f(\alpha)$ to $h(q)$,
\begin{equation} \alpha = h(q) + q h'(q) \quad {\rm and} \quad
f(\alpha) = q [\alpha - h(q)] + 1.\label{Legendre2} \end{equation}
 H\"older exponent denotes monofractality, while in
multifractal case, different parts of structure are characterized
by different values of $\alpha$, leading to existence of spectrum
$f(\alpha)$.

\begin{figure}[t]
\includegraphics[width=9cm]{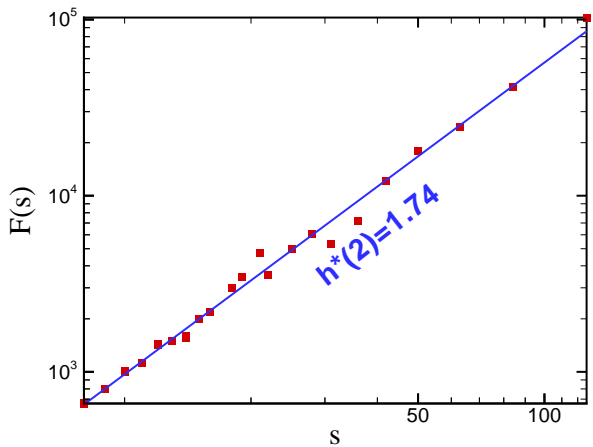}
\caption{ The log-log plot $F(s)$ versus $s$ for $q=2.0$ after
double profiling for Invention no. 1.} \label{FIG:nsvst3}
\end{figure}

\begin{figure}[t]
\includegraphics[width=9cm]{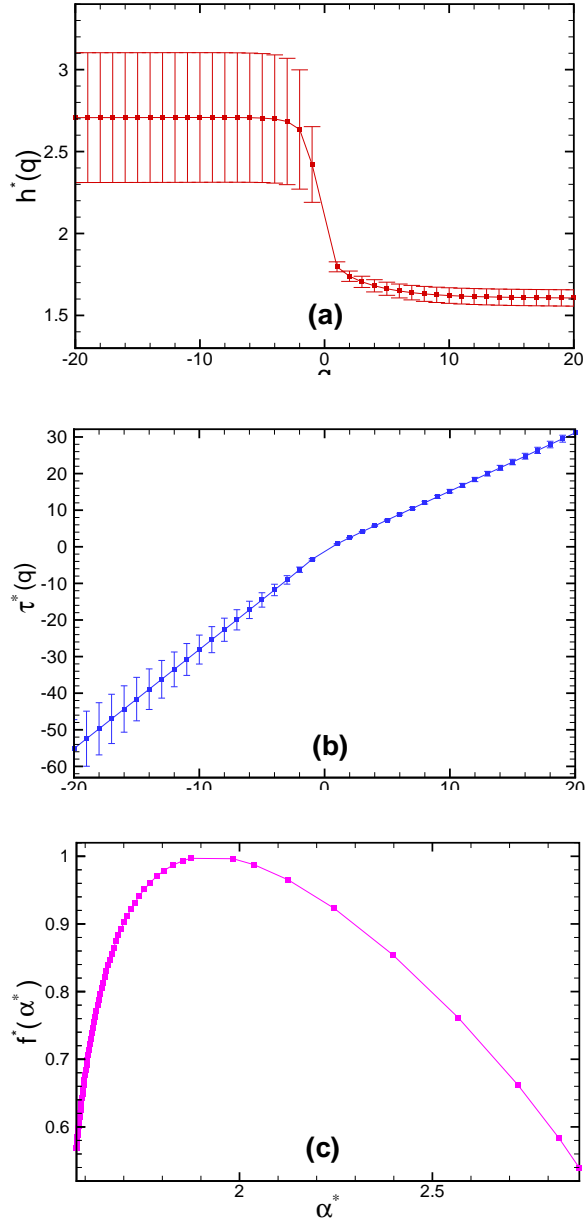}
\caption{ \small
  The q dependence of the exponents $h^*(q)$, $\tau^*(q)$ and singularity spectrum $f^*(\alpha^*)$, after
double profiling, are shown in the upper to lower panels
respectively for Invention no. 1 .} \label{fig4}\end{figure}

\begin{figure}[t]
\includegraphics[width=9cm]{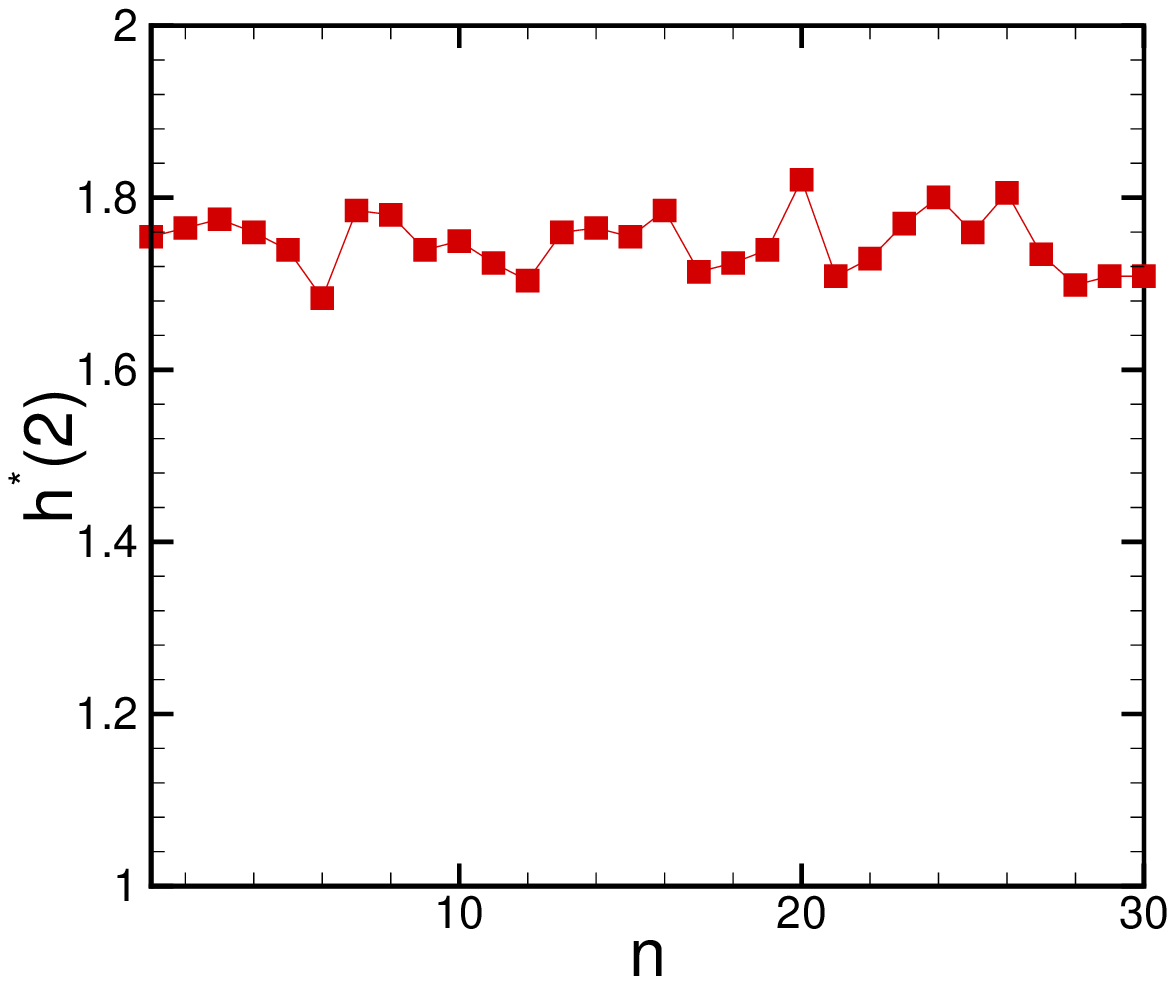}
\caption{ \small
  The second moment exponents, $h^*(q=2)$, after
double profiling, of various Inventions and Sinfonias in their
number ordering. } \label{fig5}\end{figure}

\section{Analysis of music frequency series} \label{detcross}

As mentioned in section II, a spurious of correlations may be
detected if time series is non-stationarity, so direct calculation
of correlation behavior, spectral density exponent, fractal
dimensions etc., don't give us a reliable results. It can be checked
out that, frequency series is non-stationary. One can verified
non-stationarity property experimentally by measuring stability of
average and variance in a moving window for example using scale $s$.
According to MF-DFA1 method, generalized Hurst exponents $h(q)$ in
eq.(\ref{Hq}) can be found by analyzing log-log plots of $F_q(s)$
versus $s$ for each $q$. Hurst exponent is between $0<H<1$. However,
the MF-DFA method can only determine positive generalized Hurst
exponents, in order to refine the analysis near the fGn/fBm boundary
or strongly anti-correlated signals when is close to zero. The
simplest way to analyze such data is to integrate the time series
before the MF-DFA procedure. Hence, we replace the single summation
in Eq. (1), which is describing the profile from the original data,
by a double summation by using Classification by the signal
summation conversion method (SSC). After using SSC method, fGn
switch to fBm and fBm switch to sum-fBm. In this case the relation
between the new exponent, $h^*(q=2)$, and $h(q = 2)$ is $h(q=2)=
h^*(q = 2)-1$ \cite{eke02,Movahed,bun02} (recently Movahed et al.
have proven the relation between derived exponent from double
profile of series in DFA method and $h(q = 2)$ exponent in Appendix
of \cite{Movahed}). We find $h^*(q=2)=1.74 \pm 0.03$ of Invention
no. 1 by using SSC method. Note that, the exponents of the new
series, after double profiling (SSC), are different from before SSC.
Therefore we name these new exponents as $h^*(q)$, $\tau^*(q)$ and
$f^*(\alpha^*)$.

Results using MF-DFA1 method for frequency signal are shown in
Figure \ref{fig4}, which shows that frequency series is a
multifractal process as indicated by strong $q$ dependence of the
exponents $h^*(q = 2)$ and $\tau^*(q)$\cite{bun02}. $q$ dependence
of the multifractal scaling exponent $\tau^*(q)$ has linear
behaviors for $q>0$ and $q<0$ and slopes of $\tau^*(q)$ are
$1.53\pm0.01$ and $2.61\pm0.01$, respectively. The values which
derived for quantities of MF-DFA1 method for Invention no. 1 are
given in Table I. We have calculated exponent $h^*(q)$ for other
Inventions and Sinfonias as well, and all of them are in the
$1.7-1.8$ range (Fig. \ref{fig5}). Figure .~\ref{fig4}(c) shows the
width of singularity spectrum, $f^*(\alpha^*)$, i.e. $\Delta
\alpha=\alpha^*(q_{max})-\alpha^*(q_{min})$ for the series is
approximately, $1.70$. It value shows the power of multifractality
of interevents is very strong \cite{paw05}. The values of derived
quantities from MF-DFA1 method, are given in Table I. There are some
specific lengths in music that have important effects in music's
statistical parameters. a) Rhythm, b) Scale is a set of tones each
having a definite pitch (perceived fundamental frequency of a sound)
and each having a specific frequency ratio compared to the others,
that is, each having specific interval relative to all the other
pitches, arranged in a sequence from low to high, or alternatively,
from high to low. Composers often transform musical patterns by
moving every note in the pattern by a constant number of scale
intervals. Since the intervals of a scale can have various sizes,
this process introduces subtle melodic and harmonic variation into
the music. This variation is what gives scalar music much of its
complexity.), c) Bar or measure is a segment of time defined as a
given number of beats (the basic time unit of a piece) of a given
duration. More over in the Persian music (Eastern music), there are
others characteristic lengths such as Gushe which is based on the
change of the tonic or stop notes. Existence of these characteristic
lengths can effect on music complexity. The music style and more
importantly composers can use various the characteristic  lengths in
the music. We check a few pieces of other genders of music like jazz
and Persian traditional music (Eastern music) that obtained $h^*(2)$
in ranges, $(1.75-1.9)$ and $(1.3-1.7)$, respectively. We find more
characteristic lengths in the Persian traditional music which leads
to decrease the exponents.

Usually, in MF-DFA method, deviation from a straight line in
log-log plot of eq.(\ref{Hq}) occurs for small scales $s$. This
deviation limits capability of DFA to determine correct
correlation behavior for very short scales and in  regime of small
$s$. The modified MF-DFA is defined as follows \cite{physa}:

\begin{eqnarray}
F^{\rm mod}_q(s) = F_q(s) {\langle [F_q^{\rm shuf}(s')]^2
\rangle^{1/2} \, s^{1/2} \over \langle [F_q^{\rm shuf}(s)]^2
\rangle^{1/2} \,
s'^{1/2} } \quad {\rm (for} \, s' \gg 1),\nonumber\\
\label{fmod}
\end{eqnarray}
where $\langle [F_q^{\rm shuf}(s)]^2 \rangle^{1/2}$ denotes the
usual MF-DFA fluctuation function, defined in eq.(\ref{fdef}),
averaged over several configurations of shuffled data taken from
original time series, and $s' \approx N/40$. The values of $h^*(q)$
exponent obtained by modified MF-DFA1 methods for frequency series
time series is $1.70\pm0.03$. The relative deviation of the new
exponent which is obtained by modified MF-DFA1 in comparison to
MF-DFA1 for original data is less than $5\%$.

\begin{figure}[t]
\includegraphics[width=9cm]{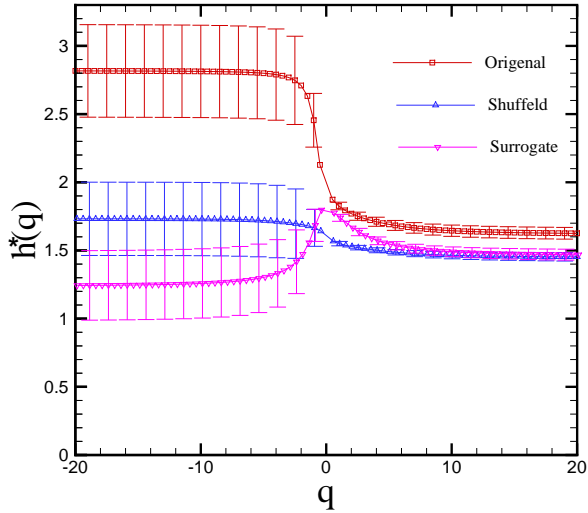}
\caption{ \small
  The new exponent, $h^*(q)$, after
double profiling, as a function of q for the original, surrogate and
shuffled data.} \label{fig6}\end{figure}

Now, we are interested in to determine  source of multifractality.
In general, two different types of multifractality in time series
can be distinguished: (i) multifractality due to a fatness of
probability density function (PDF) of the time series. In this case
multifractality cannot be removed by shuffling series. (ii)
multifractality due to different correlations in  small and large
scale fluctuations. In this case  data may have a PDF with finite
moments, e.g. a Gaussian distribution. Thus corresponding shuffled
time series will exhibit mono-fractal scaling, since all long-range
correlations are destroyed by shuffling procedure. If both kinds of
multifractality are present, shuffled series will show weaker
multifractality than original series. The easiest way to clarify the
type of multifractality, is by analyzing corresponding shuffled and
surrogate time series. Shuffling of time series destroys long-range
correlation, therefore if multifractality belongs only to long-range
correlation, we should find $h^*_{\rm shuf}(q) = 1.5$. The
multifractality nature due to fatness of
 PDF signals is not affected by shuffling procedure. On the
other hand, to determine  multifractality due to  broadness of PDF,
phase of discrete fourier transform (DFT) coefficients of frequency
series time series are replaced with a set of pseudo independent
distributed uniform $(-\pi,\pi)$ quantities in surrogate method. The
correlations in surrogate series do not change, but probability
function changes to Gaussian distribution. If multifractality in
time series is due to a broad PDF, $h^*(q)$ obtained by  surrogate
method  will be independent of $q$. If both kinds of multifractality
are present in frequency series time series, shuffled and surrogate
series will show weaker multifractality than original one.

To check  nature of multifractality, we compare  fluctuation
function $F_q(s)$, for original series (after cancelation of
sinusoidal trend) with  result from corresponding shuffled,
$F_q^{\rm shuf}(s)$ and surrogate series $F_q^{\rm sur}(s)$.
Differences between these two fluctuation functions with original
one, directly indicate  presence of long-range correlations or
broadness of probability density function in original series.
These differences can be observed in ratio  plot of $F_q(s) /
F_q^{\rm shuf}(s)$ and $F_q(s) / F_q^{\rm sur}(s)$ with respect
$s$ \cite{bun02}. Since  anomalous scaling due to a broad
probability density affects both of $F_q(s)$ and $F_q^{\rm
shuf}(s)$ in same way, only multifractality due to correlations
will be observed in $F_q(s) / F_q^{\rm shuf}(s)$. Scaling behavior
of these ratios are
\begin{equation} F_q(s) / F_q^{\rm shuf}(s) \sim s^{h^*(q)-h^*_{\rm
shuf}(q)} = s^{h^*_{\rm cor}(q)}, \label{HqCor} \end{equation}
\begin{equation} F_q(s) / F_q^{\rm sur}(s) \sim s^{h^*(q)-h^*_{\rm
sur}(q)} = s^{h^*_{\rm PDF}(q)}. \label{Hqpdf} \end{equation}
 If only fatness of PDF is responsible for
 multifractality, one
should have $h^*(q)=h^*_{\rm shuf}(q)$ and $h^*_{\rm cor}(q)=0$. On
the other hand, deviations from $h^*_{\rm cor}(q) =0$ indicates
existence of correlations, and $q$ dependence of $h^*_{\rm cor}(q)$
indicates that multifractality is due to long rage correlation. If
only correlation multifractality is present, one finds $h^*_{\rm
shuf}(q)=1.5$. If both distribution and correlation multifractality
are present, both of $h^*_{\rm shuf}(q)$ and $h^*_{\rm sur}(q)$ will
depend on $q$. The $q$ dependence of exponent $h^*(q)$ for original,
surrogate
 and shuffled time series are shown in
 Figures \ref{fig6}. $q$ dependence of $h_{\rm cor}$ and
 $h^*_{\rm PDF}$ shows that multifractality nature of frequency series time
series is due
 to both broadness of PDF and long-range correlation. Absolute value
of $h^*_{\rm PDF}(q)$ is
 greater than $h^*_{\rm cor}(q)$, so multifractality due to correlation
 is weaker than  mulifractality due to fatness.

Deviation of $h^*_{\rm sur}(q)$ and $h^*_{\rm shuf}(q)$
 from $h^*(q)$ can be determined by using $\chi^2$ test as follows:
\begin{equation}
 \chi^2_{\diamond}=\sum_{i=1}^{N}\frac{[h^*(q_i)-h^*_{\diamond}(q_i)]^2}{\sigma(q_i)
^2+\sigma_{\diamond}(q_i)^2},
 \label{khi} \end{equation}
symbol $``\diamond"$ can be replaced by $``\rm sur"$ and $``\rm
shuf"$, to determine the confidence level of $h^*_{\rm sur}$ and
$h^*_{\rm shuf}$ to the new exponents, $h^*(q)$, of original series,
respectively. Reduced chi-square
$\chi^2_{\nu\diamond}=\frac{\chi^2_{\diamond}}{\cal{N}}$ ($\cal{N}$
is number of degree of freedom) for shuffled and surrogate time
series are $16.07$, $21.30$, respectively.
 Width of singularity spectrum, $f^*(\alpha^*)$, i.e. $\Delta
\alpha=\alpha^*(q_{min})-\alpha^*(q_{max})$ for original, surrogate
and shuffled time series are approximately, $1.04$, $0.65$ and
$0.92$ respectively. These values conclude that  multifractality due
to fat tail is dominant\cite{paw05}.

Values of the exponents, $h^*(q=2.0)$, $\tau^*(q=2)$ and width of
singularity spectrum, $f^*(\alpha^*)$, $\Delta \alpha$ for the
original, shuffled and surrogate of frequency series obtained with
MF-DFA1 method are reported in Table I.

\begin{table}[t]
\begin{center}
\label{Tab} \caption{ Values of $h^*(q=2)$, $\tau^*(q=2)$ exponents
and Width of singularity spectrum, $f^*(\alpha^*)$, $\Delta \alpha$
for $q=2.0$ of Invention no.1 obtained by MF-DFA1.}
\begin{tabular}{|c|@{\hspace{0.5cm}}c@{\hspace{0.5cm}}|@{\hspace{0.5cm}}c@{\hspace{0.5cm}}|@{\hspace{0.5cm}}c@{\hspace{0.5cm}}|}
 \hline
           & $h^*(2)$ &  $\tau^*(2)$ & $\Delta\alpha$ \\ \hline
 Original  & $1.74\pm 0.03$ &$2.48\pm0.03$ &$1.70$ \\ \hline
 Surrogate &$1.71\pm0.03$ &$2.44\pm0.03$ &$0.80$  \\ \hline
 Shuffled  & $0.50\pm0.03$ &$0.02\pm0.03$ & $0.50$  \\ \hline
\end{tabular}
\end{center}
\end{table}

\section{Conclusion}

MF-DFA method allows us to determine multifractal characterization
of non-stationary and stationary time series. We have shown that
MF-DFA1 result of frequency series of Bach Inventions. Applying
MF-DFA1 method is demonstrated that frequency series have long term
correlation. We calculated the second moment exponent after the new
profile for other Inventions and Sinfonias and they are in the
$1.7-1.8$ range. $q$ dependence of $h^*(q)$ and $\tau^*(q)$, shows
that frequency series has multifractal behavior. By comparing the
second moment exponent of original time series with shuffled and
surrogate one's, we have found that multifractality due to broadness
of probability density function has more contribution than
correlation in Inventions.

\section{Acknowledgment}
We would like to thank M Sadegh Movahed for reading the manuscript
and useful comments. GRJ would like to acknowledge the hospitality
extended during his visits at the IPAM, UCLA, where this work was
started.

\bibliography{apssamp}

\begin{thebibliography} {50}

\bibitem{Noralv} http://www.ntnu.no/

\bibitem{lyan} Julyan H. E. Cartwright, Diego L. Gonz´alez, and Oreste
Piro, 1999 Phys. Rev. Lett. {\bf 82}, 5389

\bibitem{Heather} Heather D. Jennings, Plamen Ch. Ivanov,
Allan de M. Martins, P.C. da Silva, Viswanathan G M, Physica A 336
(2004) 585 – 594.

\bibitem{Jafari} Jafari G R, Pedram P, Ghafori K, AIP Proceeding (2006).

\bibitem{Gonzalez} Gonzalez D L, Morettini L, Sportolari F,
Rosso O, Cartwright J H E and Piro O, arXiv:chao-dyn/9505001,
(1995).

\bibitem{Shi} YU SHI, arXiv:adap-org/9509001, (1995).

\bibitem{Diodati} Diodati p and Piazz P, Eur. Phys. J. B {\bf 17},
143-145.

\bibitem{Jean} Jean Pierre Boon  and Olivier Decroly, 1995, Chaos 5(3)
501-508.

\bibitem{Rudi1} Rudi Cilibrasi, Paul Vitanyi and Ronald de Wolf,
Computer Music Journal, 28:4, pp. 49–67, Winter 2004.

\bibitem{Rudi2} Rudi Cilibrasi and Paul Vitanyi,
2005, IEEE TRANSACTIONS ON INFORMATION THEORY, 51(4), 1523–1545.

\bibitem{Peng94} Peng C K, Buldyrev S V, Havlin S, Simons M, Stanley H
 E, and Goldberger A L, 1994 Phys. Rev. E {\bf 49}, 1685
; Ossadnik S M, Buldyrev S B, Goldberger A L, Havlin S, Mantegna R
N, Peng C K, Simons M and Stanley H E, 1994 Biophys. J. {\bf
 67}, 64

\bibitem{hurst65} Hurst H E, Black R P and Simaika Y M, 1965
{\em Long-term storage. An experimental study}  (Constable,
London)

\bibitem{wtmm} Muzy J F, Bacry E and Arneodo A, 1991 Phys. Rev. Lett. {\bf 67},
3515

\bibitem{murad} Taqqu M S, Teverovsky V and Willinger W, 1995
 Fractals {\bf 3}, 785

\bibitem{physa} Kantelhardt J W, Koscielny-Bunde E, Rego H H A,
Havlin S and Bunde A, 2001 Physica A {\bf 295}, 441

\bibitem{kunhu} Hu K, Ivanov P Ch, Chen Z, Carpena P and Stanley H E, 2001
Phys. Rev. E {\bf 64}, 011114

\bibitem{kunhu1} Chen Z, Ivanov P Ch, Hu K and Stanley H E, 2002
 Phys. Rev. E {\bf 65}, preprint physics/041107.


\bibitem{dns} Buldyrev S V, Goldberger A L, Havlin S, Mantegna R N, Matsa M E,
Peng C K, Simons M and Stanley H E, 1995 Phys. Rev. E {\bf 51},
5084; Buldyrev S V, Dokholyan N V, Goldberger A L, Havlin S, Peng
C K, Stanley H E and Viswanathan G M, 1998 Physica A {\bf 249},
430

\bibitem{herz} Ivanov  P Ch, Bunde A, Amaral L A N, Havlin S, Fritsch-Yelle J,
Baevsky  R M, Stanley H E and Goldberger A L, 1999 Europhys. Lett.
{\bf 48}, 594; Ashkenazy Y, Lewkowicz M, Levitan J, Havlin S,
Saermark K, Moelgaard H, Thomsen P E B, Moller M, Hintze U and
Huikuri H V, 2001 Europhys. Lett. {\bf 53}, 709; Ashkenazy Y,
Ivanov P Ch, Havlin  S, Peng C K, Goldberger A L and Stanley H E,
2001 Phys. Rev. Lett. {\bf 86}, 1900

\bibitem{Peng95} Peng C K, Havlin  S, Stanley H E and Goldberger A L, 1995
Chaos {\bf 5} 82

\bibitem{PRL00} Bunde A, Havlin S, Kantelhardt J W, Penzel T, Peter J H and
Voigt K, 2000 Phys. Rev. Lett. {\bf 85}, 3736

\bibitem{neuron} Blesic S, Milosevic  S, Stratimirovic D and
Ljubisavljevic  M, 1999 Physica A {\bf 268}, 275; Bahar S,
Kantelhardt J W, Neiman A, Rego  H H A, Russell D F, Wilkens L,
Bunde A and Moss F, 2001 Europhys. Lett. {\bf 56}, 454

\bibitem{gait} Hausdorff J M, Mitchell S L, Firtion R, Peng C K, Cudkowicz M E,
Wei J Y and Goldberger A L, 1997 J. Appl.
 Physiology {\bf 82}, 262

\bibitem{Govindan} Govindon R B and Kantz H, Europhysics letters,
\textbf{68} (2), pp: 184-190 (2004)

\bibitem{wetter} Koscielny-Bunde E, Bunde A, Havlin  S, Roman H E,
Goldreich  Y and Schellnhuber H J, 1998 Phys. Rev. Lett. {\bf 81},
729; Ivanova K and Ausloos M, 1999 Physica A {\bf 274}, 349;
Talkner P and Weber R O, 2000 Phys. Rev. E {\bf 62}, 150

\bibitem{cloud} Ivanova K, Ausloos M, Clothiaux E E and Ackerman T P, 2000
Europhys. Lett. {\bf 52}, 40

\bibitem{malamudjstatlaninfer1999} Malamud B D and Turcotte D L, 1999 J. Stat.
Plan. Infer. {\bf 80}, 173

\bibitem{Alados2000} Alados C L and Huffman M A, 2000 Ethnology {\bf 106}, 105

\bibitem{economics} Mantegna R N and Stanley H E, 2000 {\it An
 Introduction to Econophysics} (Cambridge University Press,
 Cambridge); Liu Y, Gopikrishnan P, Cizeau P, Meyer M, Peng C K and Stanley H
E, 1999 Phys. Rev. E {\bf 60}, 1390; Vandewalle N, Ausloos M and
Boveroux P, 1999 Physica A {\bf 269}, 170

\bibitem{fest} Kantelhardt J W, Berkovits R, Havlin S and Bunde A, 1999 Physica
A {\bf 266}, 461; Vandewalle N, Ausloos M, Houssa M, Mertens P W
and Heyns M M, 1999 Appl. Phys. Lett. {\bf 74}, 1579

\bibitem{feder88} Feder J, 1988 {\em Fractals} (Plenum Press, New York)

\bibitem{barabasi} Barab\'asi A L and Vicsek T, 1991 Phys. Rev. A
 {\bf 44}, 2730
\bibitem{peitgen}  Peitgen H O, J\"urgens H and Saupe D, 1992 {\it
 Chaos and Fractals} (Springer-Verlag, New York), Appendix B

\bibitem{bacry01} Bacry E, Delour J and Muzy J F, 2001 Phys. Rev. E {\bf 64},
026103

\bibitem{fano} Fano U, 1947 Phys. Rev. {\bf 72} 26

\bibitem{allan} Barmes J A and Allan D W, 1996 Proc. IEEE {\bf 54} 176

\bibitem{buldy95} Buldyrev S V, Goldberger A L, Havlin  S, Mantegna R N, Matsa
M E, Peng C K, Simons M, Stanley H E, 1995 Phys. Rev. E {\bf 51}
5084

\bibitem{eke02}
Eke A, Herman P, Kocsis L and Kozak L R, 2002 Physiol. Meas. {\bf
23}, R1-R38

\bibitem{Movahed} M Sadegh Movahed, G R Jafari, F Ghasemi , Sohrab Rahvar
and M Reza Rahimi Tabar, J. Stat. Mech. (2006) P02003.

\bibitem{bun02}
Kantelhardt J W, Zschiegner S A, Kosciliny-Bunde E, Bunde A, Pavlin
S and Stanley H E, 2002 Physica A {\bf 316}, 78-114.

\bibitem{paw05} O\'{s}wi\c{e}cimka P and {\it et. al},
[arXive:cond-mat/0504608]
\end{thebibliography}

\end{document}